\documentclass[aps,prl,reprint,superscriptaddress,showpacs,showkeys,longbibliography]{revtex4-1}

\usepackage[]{hyperref} 
\hypersetup{pdfauthor={Shalaev, Walasik, Litchinitser},pdftitle={Experimental Valley-Hall effect}}
\usepackage{amsmath}
\usepackage{amsfonts}
\usepackage{textcomp}
\usepackage[]{graphicx}
\usepackage{epstopdf}
\usepackage{mathrsfs}
\usepackage{amsbsy}
\usepackage{cleveref}
\usepackage{lipsum}

\crefformat{equation}{Eq.~(#2#1#3)}
\crefformat{figure}{Fig.~#2#1#3}

\Crefformat{equation}{Equation~(#2#1#3)}
\Crefformat{figure}{Figure~#2#1#3}

\begin{document}

\title{Experimental demonstration of valley-Hall topological photonic crystal at telecommunication wavelengths}

\author{Mikhail I. Shalaev}
\affiliation{Department of Electrical Engineering, University at Buffalo, The State University of New York, Buffalo, New York 14260, USA}
\author{Wiktor Walasik}
\affiliation{Department of Electrical Engineering, University at Buffalo, The State University of New York, Buffalo, New York 14260, USA}
\author{Alexander Tsukernik}
\affiliation{Toronto Nanofabrication Centre, University of Toronto, Toronto, ON M5S 3G4, Canada}
\author{Yun Xu}
\affiliation{Department of Electrical Engineering, University at Buffalo, The State University of New York, Buffalo, New York 14260, USA}
\author{Natalia M. Litchinitser}
\affiliation{Department of Electrical Engineering, University at Buffalo, The State University of New York, Buffalo, New York 14260, USA}

\date{\today}

\begin{abstract}
Photonic topological insulators provide unprecedented possibilities to eliminate scattering losses and improve the efficiency of optical communication systems. 
Despite significant theoretical efforts, the experimental demonstration of an integrated photonic topological insulator operating in the telecommunication regime was still missing. 
Here, we design, fabricate and characterize a photonic-crystal-based topological structure that exhibits valley-Hall effect. 
We experimentally demonstrate the  propagation of topologically protected edge states in a CMOS-compatible chip operating at telecommunication wavelengths.
This contribution is an important step towards integrated topological photonics.

\end{abstract}

\maketitle
\vspace{-2em}

Discovery of topological insulators (TIs)---materials that insulate in the bulk but allow conduction at the surface---led to a breakthrough in solid-state physics~\cite{kane05a,kane05,moore10,Bernevig13,fereira13,katmis16}. 
The conducting properties of TIs are defined by the spatio-temporal symmetries of their structure.
Breaking of the time-reversal symmetry may lead to the quantum-Hall effect that features unidirectional transport at the material surfaces due to the absence of the counter-propagating states.
Breaking of certain spatial symmetries leads to the rise of spin- or valley-Hall effects, in this case the two counter-propagating edge states have opposite circular polarizations (spins) or belong to a different regions (valleys) in the reciprocal space. 
As a result, the coupling between these states is reduced, and the transport is protected against backscattering on defects that do not induce inter-valley scattering~\cite{Ju15}.

Recently, the concept of TIs has been extended to photonics~\cite{Khanikaev12,lu14,Khanikaev17}. 
Various approaches to construct TIs for light have been proposed.
Photonic crystals (PCs)~\cite{jannopoulos08} incorporating gyromagnetic materials have been shown to exhibit an analogue of the quantum-Hall effect in the presence of external magnetic fields~\cite{haldane08,wang08,wang09,poo11}.
Arrays of coupled ring-resonators~\cite{hafezi13,hafezi11,fang12} or helical waveguides~\cite{Rechtsman13} allow one to emulate the time-reversal-symmetry breaking and lead to unidirectional transport in edge-states.
Photonic analogues of spin- and valley-Hall effects that require breaking of spatial instead of temporal symmetries have been demonstrated in PCs made of metallic~\cite{chen14,ma15,Ma17,Wu17} or dielectric~\cite{Ma16,Chen17,Dong17,wu15} materials.
Despite substantial theoretical effort, topologically protected light propagation in PC structures was experimentally confirmed only in the GHz regime~\cite{chen14,hafezi11,hafezi13,cheng16,Wu17,poo11,wang09,slobozhanyuk16}.
Quantum-Hall~\cite{Rechtsman13} and valley-Hall~\cite{Noh17} effects have been experimentally observed in visible or infre-red regimes in arrays of coupled waveguides. 
Valley-Hall effect was recently observed also in sonic crystals~\cite{Lu16}. 
Several different designs of PC lattices supporting valley-Hall effect were proposed~\cite{Chen17,Dong17,Ma16,Ma17,Ni17,Wu17}, but none of them was easily adaptable for fabrication of the device operating in the telecommunication regime.

In this letter, we experimentally demonstrate a silicon-based design of a valley-Hall TI implemented in a PC platform that can be integrated on a chip using a standard CMOS-compatible techniques and operates at telecommunication wavelengths.
We prove that the PC slab proposed here supports topologically protected light propagation and features scattering-free energy transport around sharp turns along the path.

\begin{figure*}
 	\includegraphics[width = 0.66\textwidth, clip = true, trim = {0 0 0 0}]{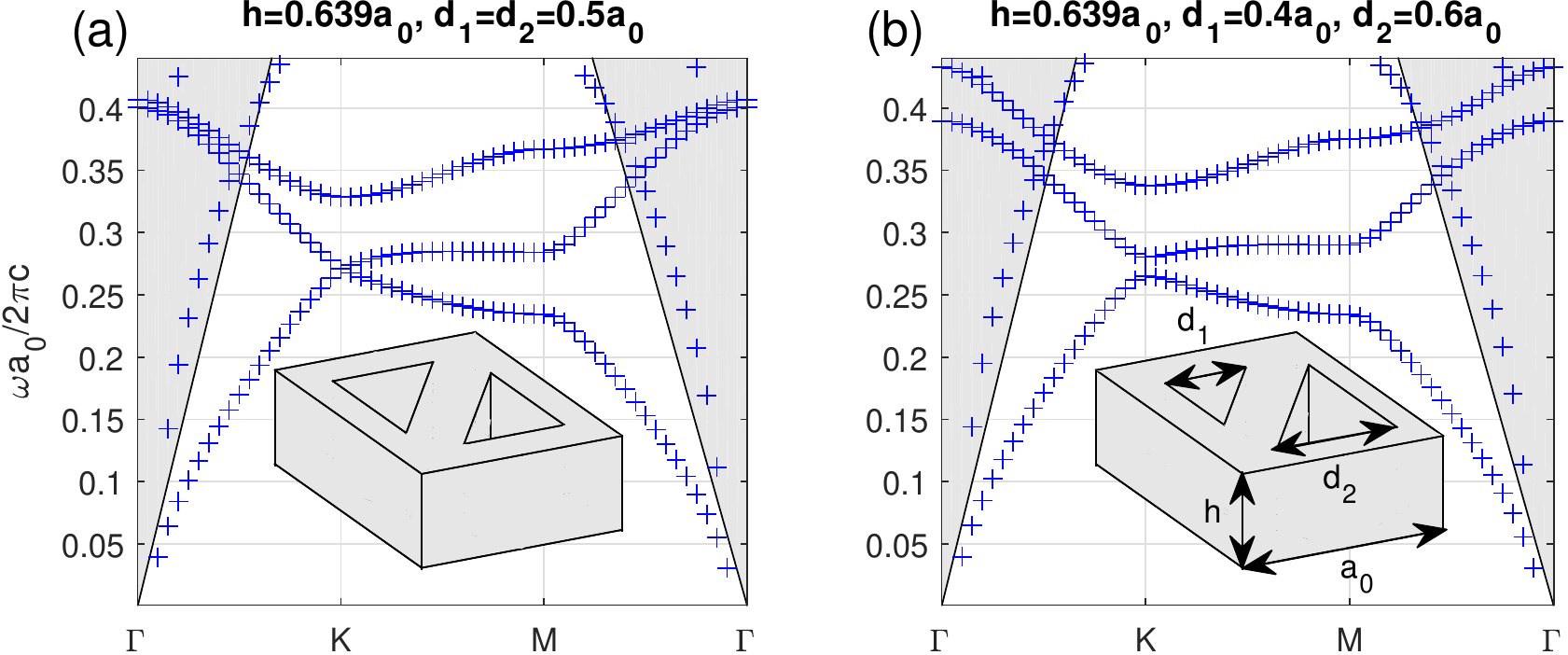}
 	\includegraphics[width = 0.32\textwidth, clip = true, trim = {0 0 0 0}]{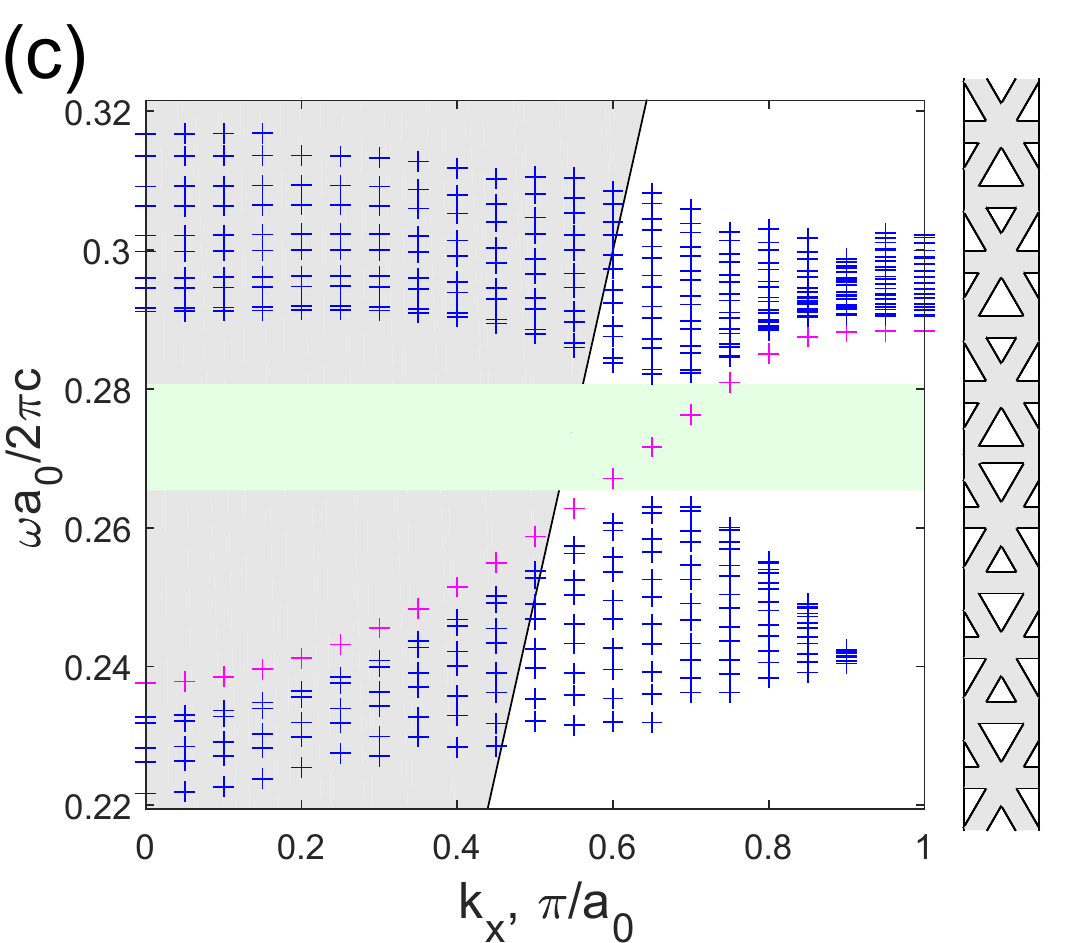}
 	\caption{
 		Band structures for the photonic-crystal slab with the $C_6$ symmetry (a) and with the $C_3$ symmetry that exhibits the valley-Hall effect (b). 
 		The insets illustrate the unit cell of the photonic-crystal slab and show the dimensions of the structure. 
 		(c)~Edge state of the periodic 3D structure illustrated in the inset. 
 		The structure is periodic along the horizontal direction and finite along the vertical direction.
 		Frequencies above the light cone are shaded in gray. 
 		The refractive index of the silicon is taken as $n_{\textrm{Si}}=3.48$.}
 	\label{fig:3Dstr}
 \end{figure*}

The three dimensional (3D) structure under investigation is shown in the insets of \cref{fig:3Dstr}, and consists of a silicon membrane with triangular holes, suspended in the air. 
The starting point of our design is a PC lattice with a $C_{\textrm{6}}$ symmetry, whose unit cell shown in the inset of \cref{fig:3Dstr}(a).
The structure features a honeycomb lattice with two inverted triangular air holes per unit cell.
The silicon PC slab was fabricated on a standard silicon-on-insulator (SOI) wafer, the silica underneath of the PC has been removed with wet etching.
This results in a symmetric air--silicon--air structure and prevents coupling between the transverse-electric-like (TE-like) and transverse-magnetic-like modes~\cite{jannopoulos08}.
 
For the lattice with equal hole sizes $d_1 = d_2$, the dispersion relation (computed with COMSOL Multiphysics~\cite{comsol}) for the TE-like polarized light propagating in the plane of the membrane features a Dirac cone at the $K$ ($K'$) points in the momentum space, as shown in \cref{fig:3Dstr}(a).
Introducing asymmetry between the sizes of the two holes withing the unit cell ($d_1 \neq d_2$) breaks the inversion symmetry and reduces the lattice symmetry to $C_3$.
As a result, the degeneracy at the $K$ ($K'$) points is lifted and a band gap opens, as shown in \cref{fig:3Dstr}(b).
A similar behavior of the band structure was recently observed in a photonic boron-nitride lattice built of circular instead of triangular holes~\cite{Collins16}.
The advantage of our PC design with triangular holes over the other proposed geometries (e.g.~with circular holes) is evident from the behavior of the band dispersion. 
In the structure with circular holes, the frequency of the second band decreases between the $K$ and the $M$ points in the reciprocal space leading to a smaller and indirect band gap.
On the contrary, the use of triangular holes allows for elevating the second band at the $M$ point resulting in appearance of a well-defined Dirac cone at the $K$ point for the case with equal air holes, and in increased band gap for dissimilar hole sizes~\cite{barik16}.
This is especially important for PCs with small holes asymmetry and in case of low refractive-index contrast between the slab and the surrounding medium.

The parameters of the structure such as, lattice constant $a_0$, the hole sizes $d_1$ and $d_2$, and the membrane height $h$, are optimized to ensure that: 
(i) the structure operates in the telecommunication wavelength range, 
(ii) the effective index of the mode is high enough to ensure that the Dirac cone is well defined or that the band gap is open,
(iii) the effective index is not too high so that the higher-order (in the out-of-plane direction) modes of the membrane are absent, eliminating the associated losses.

For the valley-Hall effect, one expects to see an edge state at an interface between the structures with two different polarities (determined in our case by the orientation of the large triangle).
\Cref{fig:3Dstr}(c) shows the band structure for a PC with such an interface.
In the upper half of the PC structure, the large triangles are pointing upwards while in the lower half of the PC, the large triangles are pointing downwards. 
The structure illustrated in the inset of \cref{fig:3Dstr}(c) is periodic in the horizontal direction, finite in the vertical direction (20 unit cells for each region), and has a thickness $h=0.639a_0$.  
The band structure reveals the presence of an edge state (magenta) crossing the band gap (shaded in green) between the first and the second bulk bands (blue). 
As the band structure is symmetric with respect to the wave vector $k_x$, there is an edge state with opposite group velocity in the second valley. 
The structure parameters ($h$, $d_1=0.4a_0$, and $d_2 = 0.6a_0$) are chosen in such a way that for the frequencies inside the band gap, the edge state is located below the light cone, reducing the losses due to light leakage~\cite{jannopoulos08}. 

\begin{figure*}
	\includegraphics[width = \textwidth, clip = true, trim = {0 0 0 0}]{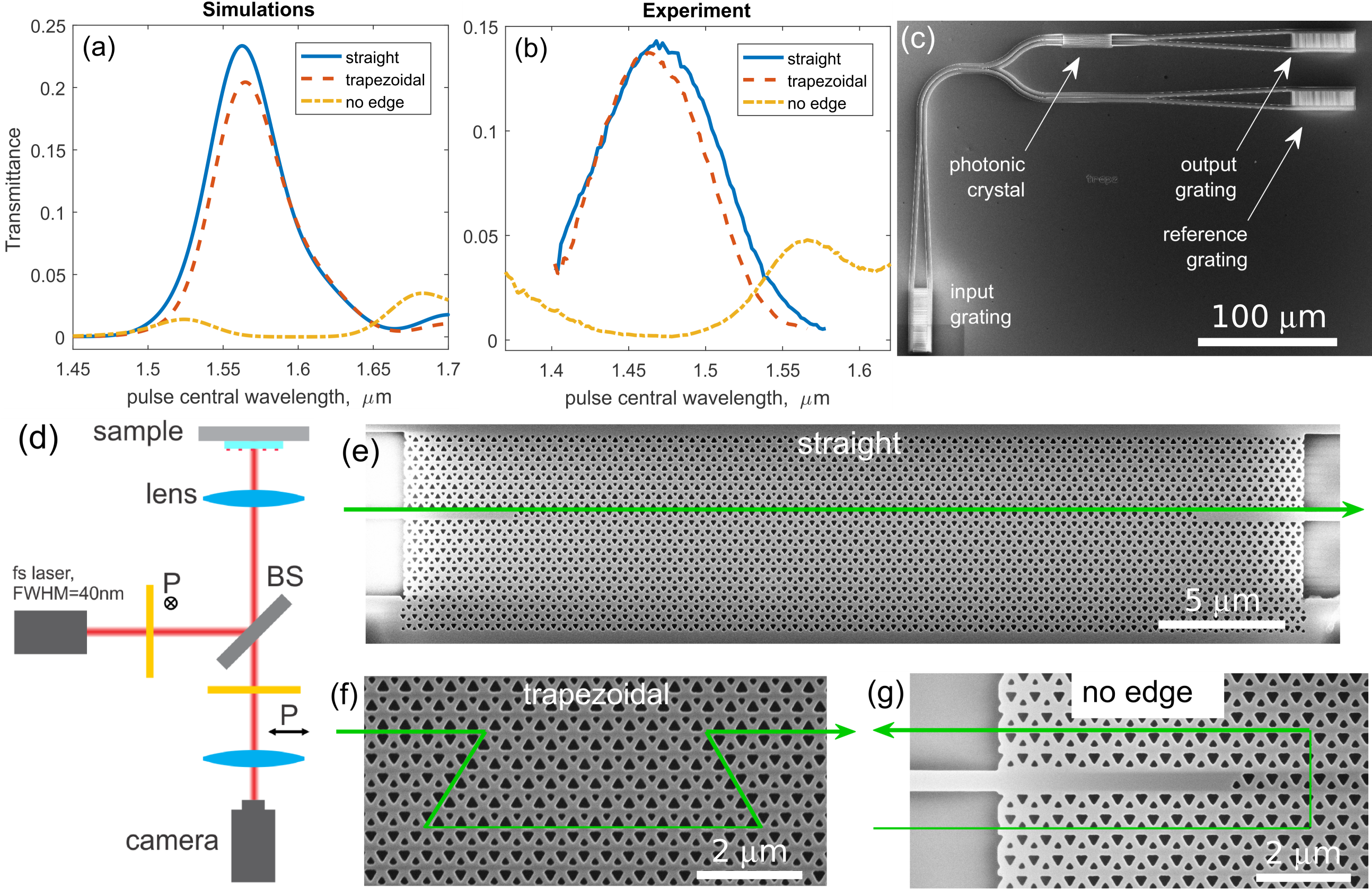}
	\caption{
		(a) Numerically computed and (b) experimentally measured transmittances for the structure with PC having single polarity without edge (orange dot-dash line), shown in (g); the structure with the straight interface (blue solid line), shown in (e); and the structure with the trapezoidal interface, shown in (f).
		(c) Scanning electron microscope image of a fabricated sample. 
		(d) Experimental setup used for transmittance measurements.}
	\label{fig:exp}
\end{figure*}

In order to confirm the valley-Hall topological protection for the edge state shown in \cref{fig:3Dstr}(c), we first, perform
analysis of Berry curvature and topology of the band structure, and then examine the propagation of light along the interface between the PCs with opposite polarities.
Later, we fabricate the structure and experimentally study its transmission properties. 
For numerical simulations, we use an effective index approximation to reduce the 3D problem to two dimensions (2D)~\cite{Qiu02,Hammer09}, and to minimize the computational cost.
As discussed above, the valley-Hall topological properties of a PC lattice stem from the broken inversion symmetry of the structure.
As the reduction of the dimensionality of the problem does not change the PC-lattice symmetry, in the following we assume that it also does not modify the topological properties of the structure.
This assumption will be confirmed by the numerical and experimental results presented in \cref{fig:exp}.

\begin{figure*}
	\includegraphics[width = \textwidth, clip = true, trim = {0 0 50 0}]{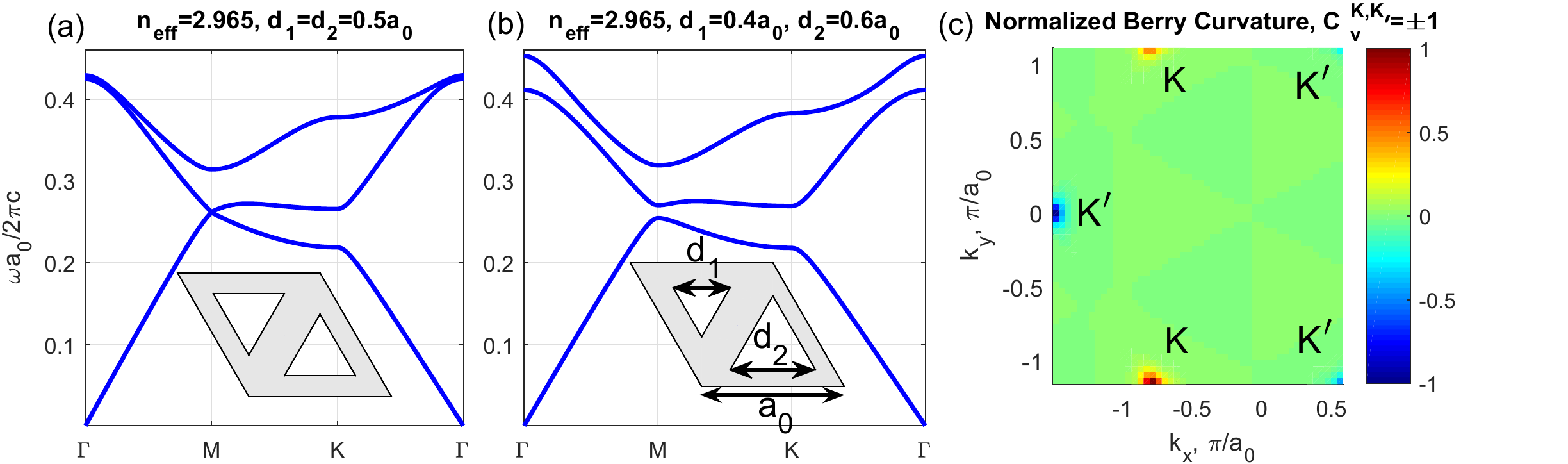}
	\caption{
		 Band structure of the 2D photonic crystal with $C_6$ symmetry (a) and the structure with broken $C_6$ symmetry (only $C_3$ symmetry is preserved) (b). 
		The schematics of the corresponding unit cells of a 2D photonic crystal are shown in the insets. 
		(c) Normalized Berry curvature of the first band of the $C_3$-symmetric photonic-crystal lattice with the valley Chern number equal to $C_v^{K/K'} = \pm 1$ corresponding to the $K$/$K'$ valleys.}
	\label{fig:2Dstr}
\end{figure*}

\begin{figure*}
	\includegraphics[width = \textwidth, clip = true, trim = {0 0 0 0}]{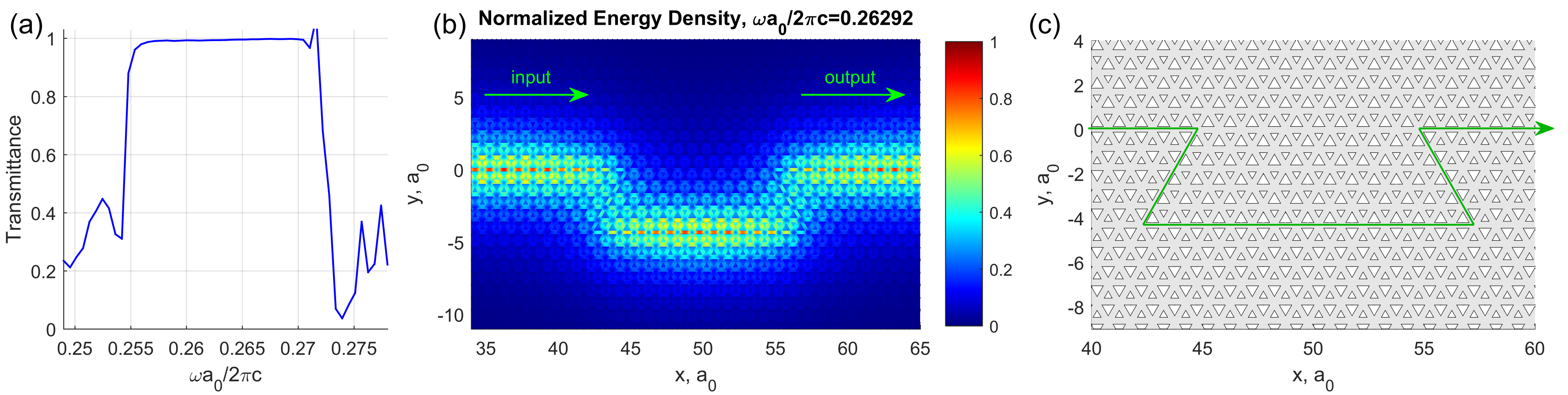}
	\caption{(a)~Transmittance spectrum for 2D photonic crystal with trapezoidally shaped edge shown in (c). 
		(b)~Normalized energy density of the propagating edge state, for the frequency located in the band gap.}
	\label{fig:3Deff}
\end{figure*} 

The insets in \cref{fig:2Dstr}(a) and (b) show the unit cells of the 2D-PC-lattice counterparts of the 3D structures studied in \cref{fig:3Dstr}(a) and (b), respectively. 
The refractive index of the silicon (shaded in gray) is assumed to be the effective index of the mode of the uniform unpatterned silicon membrane with thickness chosen above. 
Using Lumerical varFDTD software~\cite{lumerical}, we have computed the effective-index value to be $n_{\textrm{eff}} = 2.965$. 
This assumption results in a small red-shift of the position of the Dirac cone and the band gap in the 2D PC compared to the original 3D structure.
The 2D structure is assumed to be invariant along the out-of-plane direction, and the TE-polarized light propagates in-plane.
Similar to the case of the 3D structure, the geometrical modification of the hole sizes leads to the opening of the band gap  at the $K$/$K'$ points in the reciprocal space.
As a result, the band structure can be characterized by a non-zero Berry curvature, with opposite signs at the $K$ and $K'$ points. 
The distribution of the normalized Berry curvature in the first Brillouin zone for the lowest energy band is shown in \cref{fig:2Dstr}(c).
The corresponding valley Chern numbers~\cite{Fukui05,Bleu17} are calculated to be $C_v^{K/K'} = \pm 1$.

To confirm that the 2D structure possesses topological properties, we study the propagation of an edge state along the interface with sharp turns. 
We consider a trapezoidally shaped interface between the PC with opposite polarities shown in \cref{fig:3Deff}(c). 
The resulting transmittance of the TE-polarized light is illustrated in \cref{fig:3Deff}(a) and clearly reveals a region with a unitary transmittance. 
This result proves that the propagation of light along the interface between the crystals with opposite polarities is topologically protected against scattering on sharp turns. 
A typical normalized energy density distribution along the interface is shown in \cref{fig:3Deff}(b).

Next, we have fabricated the samples and measured them.
The Scanning Electron Microscope (SEM) image of the PC sample is shown in \cref{fig:exp}(c).
The structure is 82 units cells~(u.c.) long and 21~u.c.~wide, with following parameters of the unit cell: $a_0=423$~nm, $h=270$~nm, $d_1=169$~nm, $d_2 = 254$~nm.
We have used a standard fabrication techniques with the procedure similar to the one described in Ref.~\cite{fab_proc}.
Firstly, top silicon layer of the SOI was patterned by means of Electron Beam Lithography (EBL) with ZEP520A resist followed by Reactive Ion Etching (RIE) in the CF$_6$ and C$_4$F$_8$ gases.
Secondly, we used photo-lithography with S1818 photo resist in order to define a window above the PC structure.
Then selective wet etching in buffered hydrofluoric acid (BHF) was performed to remove buried oxide under the PC.

In the experiment, light is focused from free space onto an input grating followed by taper to efficiently couple light to a single-mode silicon wire waveguide.
Then energy is split using a 50/50 Y-splitter and half of the light is coupled to the PC, whereas the other half is transmitted through the silicon waveguide and used as a reference.
Our numerical simulations have shown that the transverse profile of the edge state is asymmetric with respect to the interface.
Therefore, the edge state is excited by the waveguide positioned slightly off-center with respect to the interface between the two PC (see \cref{fig:exp}(e) or (g)), which allows for efficient excitation of this asymmetric mode. 
The coupling efficiency is estimated to be above~70\%.
The light transmitted through the PC and the reference light are out-coupled from the chip using another diffraction gratings and are captured by the infra-red camera.
We designed the structure such that the input and output light have orthogonal polarizations which allowed us to distinguish the output and filter out all undesired input wave reflected from the sample. 
The transmittance of the PC structure shown in \cref{fig:exp}(b) is calculated as a ratio between the light transmitted through the PC and the reference. 
Therefore, the transmittance takes into account also all the coupling and scattering losses in the PC structure. 

The input light used in the experiment comes from a femtosecond 1~kHz laser system (Coherent Libra) coupled to an optical parametric amplifier (Light Conversion TOPAS-C), resulting in a pulsed input with the spectral full-width half-maximum (FWHM) of around 40~nm at telecommunication wavelengths.
The short pulse length (less than the length of the PC structure) allows us to avoid effects associated with Fabry-P\'erot resonances caused by reflections from the gratings and terminations of the PC structure.
The laser source used in the experiment generates pulses at a 1~$\mu$s intervals, which ensures that only one femtosecond pulse at a time propagates through the sample, preventing interference effects resulting from an overlap between the two neighboring pulses. 
Pulsed light from a femtosecond laser system is focused on the sample using an infinity corrected objective lens. 
Observation of the light out-coupled from the sample is made possible by the use of a pellicle beamsplitter (BS). 
The polarizer that transmits only the output-light polarization is used to eliminate the unwanted light with the orthogonal polarization reflected from the sample.  
The sample is imaged on the sensor of the camera using an achromatic lens, which together with the objective lens builds a 4-f system.
The use of the achromatic lens, the microscope objective, and the pellicle BS allows for reliable measurements in a wide range of frequencies reducing the chromatic aberrations.

We have performed the experiments on three different types of samples: 
(i) the sample with only one polarity of the crystal (no-edge), 
(ii) the sample with a straight interface between the crystal with opposite polarities (no sharp edges), and 
(iii) the sample with trapezoidal shape of the interface. 
The central frequency of the source was swept in order to collect the spectral dependence of the transmittance.
We observe that in the sample with no edge, there is a region of low transmittance (for wavelength range 1420--1510~nm), revealing the presence of the band gap. 
The spectral measurements of the samples with an interface, show high transmittance in the band-gap region, proving the existence of the edge state.
The transmittances of the samples with the straight and the trapezoidal interfaces are very close to each other. 
This confirms the theoretical prediction that the propagating edge state is immune to scattering at the sharp turns and proves the topological protection of the excited edge states~\cite{chen14}.
The absolute values of the transmittance of the samples are relatively low, due to coupling losses and because of the small width of the PC structure. 
The total width of the fabricated PC was 21~u.c.~and, as a result, the interface was located only 8~u.c.~away from the PC termination. 
As revealed by our numerical simulations, a portion of light traveling in an edge state is out-coupled from the PC, due to the insufficient size of the PC structure with respect to the transverse localization of the edge state. 
The size of the PC structure was limited in order to avoid bending of the suspended membrane that might occur during the drying process after etching of silicon dioxide layer.

The numerical experiments was performed using a commercial-grade simulator based on the finite-difference time-domain (FDTD) method~\cite{lumerical}. 
The propagation of light was simulated in three types of structures considered in the experiment, shown in Figs.~\ref{fig:exp}(e)--(g). 
The structure size used in the simulations was the same as on the fabricated samples.
The results of the numerical simulations are presented in \cref{fig:exp}(a) and are in good agreement with the experimental results. 
The numerical simulations were performed using the 2D FDTD method with the effective refractive index value being used for silicon. 
This approximation results in a slight shift of the spectral position of the band-gap and the high transmission region associated with the edge-state.

To conclude, we have designed and experimentally demonstrated a silicon-based CMOS compatible structure that exhibits valley-Hall effect at telecommunication wavelengths.
The structure consists of a silicon photonic-crystal membrane suspend in the air. 
The membrane is patterned in order to get photonic equivalent of a boron-nitride crystal lattice, whose unit cell contains two inverted triangular holes of a different size. 
We have fabricated the structure and characterized the topologically protected transmission along the interface between the two crystal lattices with opposite polarities. 
Our experimental results are in good agreement with the numerical simulations. 
To the best of our knowledge, this is the first experimental demonstration of a valley-Hall topologically protected transport in photonic-crystal-based structures in the telecommunication regime. 
This work is likely to be an important step towards integrated topological photonics.

\section*{Acknowledgements} 

This work was supported by Army Research Office (ARO) grants: W911NF-15-1-0152 and W911NF-11-1-0297.

The authors acknowledge fruitful discussions with professor Alexander Khanikaev from The City College of New York.


%

\end{document}